\documentclass[9pt,twocolumn,twoside,lineno]{pnas-new}

\usepackage{lipsum}
\usepackage{graphicx}   
\usepackage{color}

\usepackage{ulem}

\templatetype{pnasinvited} 

\begin{document}

\title{Generative artificial intelligence for computational chemistry: a roadmap to predicting emergent phenomena}

\author[a,b,1]{Pratyush Tiwary}
\author[b,c,2]{Lukas Herron}
\author[d,2]{Richard John}
\author[b,c,2]{Suemin Lee}
\author[a,2]{Disha Sanwal}
\author[a,2]{Ruiyu Wang}

\affil[a]{Department of Chemistry and Biochemistry and Institute for Physical Science and Technology, University of Maryland, College Park 20742, USA.}
\affil[b]{University of Maryland Institute for Health Computing, Bethesda 20852, USA.}
\affil[c]{Biophysics Program and Institute for Physical Science and Technology, 
University of Maryland, College Park 20742, USA.}
\affil[d]{Department of Physics and Institute for Physical Science and Technology, University of Maryland, College Park 20742, USA.}
\leadauthor{Tiwary}

\authorcontributions{P.T., L.H., R.J., S.L., D.S. and R.W. wrote the paper.}
\equalauthors{\textsuperscript{2}Authors two, three, four, five and six contributed equally to this work.}
\correspondingauthor{\textsuperscript{1}To whom correspondence should be addressed. E-mail: ptiwary@umd.edu}


\begin{abstract}
The recent surge in Generative Artificial Intelligence (AI) has introduced exciting possibilities for computational chemistry. Generative AI methods have made significant progress in sampling molecular structures across chemical species, developing force fields, and speeding up simulations. This Perspective offers a structured overview, beginning with the fundamental theoretical concepts in both Generative AI and computational chemistry. It then covers widely used Generative AI methods, including autoencoders, generative adversarial networks, reinforcement learning, flow models and language models, and highlights their selected applications in diverse areas including force field development, and protein/RNA structure prediction. A key focus is on the challenges these methods face before they become truly predictive, particularly in predicting emergent chemical phenomena. We believe that the ultimate goal of a simulation method or theory is to predict phenomena not seen before, and that Generative AI should be subject to these same standards before it is deemed useful for chemistry. We suggest that to overcome these challenges, future AI models need to integrate core chemical principles, especially from statistical mechanics. 
\end{abstract}

\dates{This manuscript was compiled on \today}
\doi{\url{www.pnas.org/cgi/doi/10.1073/pnas.XXXXXXXXXX}}

\maketitle
\thispagestyle{firststyle}
\ifthenelse{\boolean{shortarticle}}{\ifthenelse{\boolean{singlecolumn}}{\abscontentformatted}{\abscontent}}{}

\noindent The last few years have seen a surge of excitement and explosion of Generative Artificial Intelligence (AI) methods across scientific fields, with computational chemistry being no exception. Pioneering efforts include sampling structures and thermal distributions of complex molecular systems, developing transferable force fields, and performing accelerated simulations\cite{anstine2023generative,du2024machine,rotskoff2024sampling,mehdi2024enhanced}. With numerous tools emerging, a clear Perspective is now essential to highlight progress and critically examine pitfalls. While the scope of Generative AI's impact in chemistry is broad, this Perspective will focus exclusively on molecular simulation driven computational chemistry.  Molecular simulations offer an efficient platform for validating and iterating on new Generative AI techniques, continually improving them using force field-based approximations of reality. Additionally, using Generative AI on molecular simulation data can explore chemical and physical spaces that are otherwise difficult to access.

This Perspective is structured as follows. We begin with \textbf{The Theoretical Minimum}, summarizing the essential theoretical concepts and terminology of both Generative AI and computational chemistry. 
Next, in \textbf{Generative AI Methods for Computational Chemistry}, we provide an overview of widely used Generative AI methods  (Fig. \ref{fig:genAI_framework}), including autoencoders and their derivatives, generative adversarial networks (GANs), reinforcement learning, flow-based methods, and recurrent neural networks and language models. The Perspective then highlights \textbf{Selected applications} out of very many, focusing on \textit{ab initio} quantum chemistry, coarse-grained force fields, protein structure prediction and RNA structure prediction. Following this, the section \textbf{Desirables from Generative AI for Chemistry} explores common themes and characteristics of Generative AI tools that are particularly desirable for chemistry applications. Of particular note here is predicting emergent phenomena, which lies at the heart of chemistry and all science. Emergent phenomena occur when new properties arise even in systems with simple underlying interactions if they are large enough and/or studied long enough, as Phil Anderson discussed in his classic essay "More Is Different"\cite{anderson1972more}. Current AI approaches struggle with capturing these emergent behaviors. Recent literature on some of the most powerful Generative AI frameworks such as large language models\cite{schaeffer2024emergent} and diffusion models\cite{biroli2023generative} has highlighted and quantified the limitations of current AI tools in capturing any emergent behavior, often showing that these tools primarily excel at impressive but essentially memorization and interpolation. We conclude with brief \textbf{Critical Assessment and Outlook}, providing an honest evaluation of the progress so far and challenges that need to be addressed before Generative AI becomes a reliable member of the molecular simulator's toolbox for predicting emergent phenomena. 


\section{The Theoretical Minimum}  
In this section we will summarize the theoretical concepts and terms key to this Perspective. We do this for computational chemistry and for Generative AI. We recommend Ref. \cite{frenkel2023understanding} for a deeper understanding of computational chemistry concepts. For Generative AI there is no one book that can stay up-to-date with the dizzying pace of development, though Ref. \cite{white2021deep} covers several key underlying concepts. 
\subsection{Computational chemistry}
\label{sec:theory_compchem}
\begin{enumerate}
\item \textit{Potential energy surface}: The potential energy surface (PES) is a multidimensional surface representing the energy of a molecular system as a function of its atomic positions. The minima on the PES correspond to stable molecular structures, while the pathways connecting these minima represent possible reaction mechanisms ignoring entropic effects relevant at non-0 temperatures. 
\item \textit{Force fields}: Force fields are mathematical models used to describe the PES. They consist of a set of parameters and equations that define the interactions between atoms, including bond stretching, angle bending, torsional angles, and non-bonded interactions (e.g., van der Waals forces, electrostatic interactions). The choice of force field significantly influences the accuracy of molecular simulations, as it dictates how well the model can replicate real physical behaviors. Popular force fields include AMBER, CHARMM, OPLS and in recent years, also machine learning force fields (MLFF)\cite{nerenberg2018new}.
\item \textit{Thermodynamic ensemble}: A thermodynamic ensemble is a statistical representation of a system in which all possible microstates are considered according to specific environmental constraints like temperature, pressure, and volume. The choice of  ensemble is crucial for accurately modeling real-world conditions and predicting system behavior from molecular simulation or Generative AI.
\item \textit{Collective variables and reaction coordinate}:  
Collective variables (CVs) simplify molecular analysis by reducing dimensionality and capturing essential degrees of freedom. They are key in enhanced sampling techniques for exploring rare events. Choosing the right CVs, which approximate the reaction coordinate (RC), is vital for capturing system dynamics. The RC tracks a system’s progress along a reaction pathway, often identifying the transition state. Often the committor is considered the ideal RC\cite{best2005reaction}, as it quantifies the probability of a system evolving toward a specific product state, accounting for both energetic and entropic effects.
\item \textit{Free energy surface}: Free energy surfaces (FES) extend the concept of potential energy surfaces. They quantify the probability of observing the system as a function of one or more CVs, by marginalizing out all other degrees of freedom. Depending on how closely the chosen CVs approximate the true RC, the FES can unfortunately be mechanistically quite misleading it could be masking out true barriers. 
\item \textit{Molecular simulations}: Molecular Dynamics (MD) and Monte Carlo (MC) are essential for simulating molecular systems. MD solves Newton's equations to model atomic trajectories, while MC uses random sampling to explore configurational space. \textit{Ab initio} MD combines quantum mechanical calculations like Density Functional Theory (DFT) with MD for greater accuracy but at higher computational cost.
\end{enumerate}

\begin{figure*}[tbhp]
\centering
\includegraphics[width=.77\linewidth]{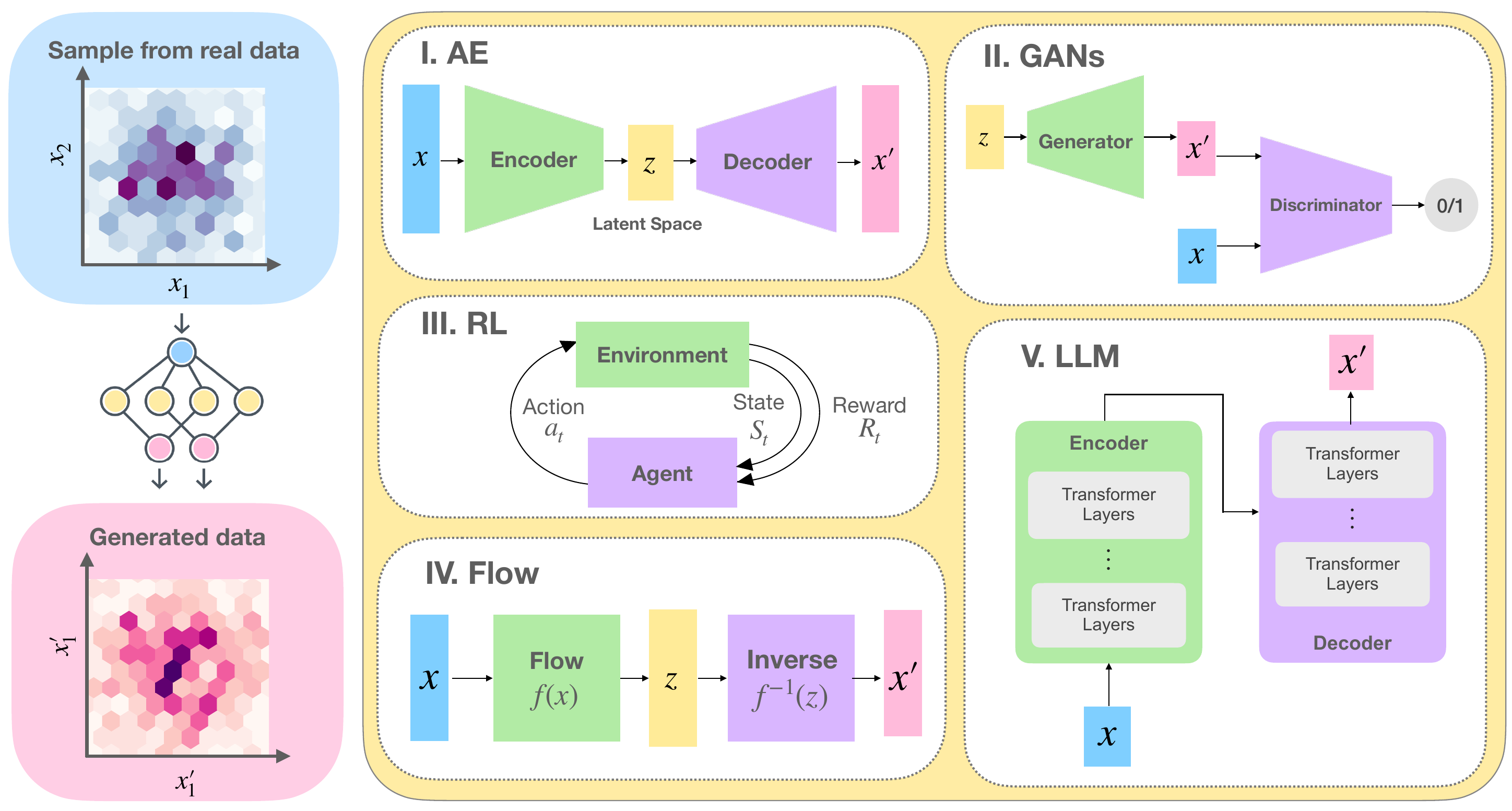}
\caption{\textbf{Overall framework of Generative AI and the methods discussed in this Perspective.} The central task in Generative AI is to generate new data that is similar to the training data or to model the underlying distribution of the data when the probability distribution is not explicitly available. This challenge is particularly relevant chemistry, where data can be structured (e.g., molecular graphs) or unstructured, time series or static. The methods discussed in this Perspective include \textbf{(I.)} Autoencoder (AE): An architecture where the encoder compresses the input data \( x \) into a latent space \( z \), and the decoder reconstructs the data from \( z \) to produce an output \( x' \) which should closely match the input. \textbf{(II.)} Generative Adversarial Networks (GANs): A framework comprising a generator that produces synthetic data \( x' \) from latent variables \( z \) and a discriminator that distinguishes between real data \( x \) and generated data \( x' \). The generator and discriminator are trained together in an adversarial process, with the generator improving its ability to create realistic data as the discriminator refines its ability to detect fakes. \textbf{(III.)} Reinforcement Learning (RL): A learning paradigm where an agent, typically a decision-making entity, interacts with an environment over time \( t \). The agent takes actions \( a_t \) based on the current state \( S_t \) of the environment, and in return, it receives rewards \( R_t \). Through this process, the agent learns to maximize cumulative rewards by refining its strategy or policy over successive iterations. \textbf{(IV.)} Flow models: These models learn to transform complex probability distributions of the data \( x \) into simpler, tractable prior distributions \( z \) using invertible functions \( f(x) \). Given data from a complex true distribution, these models enable the mapping to a simpler latent space, from which new data can be generated by inverting the transformation. \textbf{(V.)} Large Language Models (LLMs): A typical LLM consists of an encoder and a decoder, both composed of multiple transformer layers. These layers use self-attention mechanisms to understand and focus on the most relevant parts of the input sequence, facilitating the generation of coherent and contextually appropriate outputs in a variety of natural language processing tasks.}
\label{fig:genAI_framework}
\end{figure*}

\subsection{Generative AI}
\label{sec:theory_genAI}

\begin{enumerate}
\item \textit{Latent Variables}: Latent variables are hidden factors that capture underlying structures in data. In models like autoencoders and pre-deep learning methods such as Principal Component Analyses, these variables represent the reduced-dimensional space that captures essential features of the data, facilitating generation and reconstruction processes. Generally latent variables are often entangled, meaning they mix multiple underlying factors. Disentangling them improves interpretability and control, enabling precise manipulation of control parameters.
\item \textit{Prior}: The prior is a distribution over the latent variables that encodes initial beliefs about their values before observing data. The use of priors can help achieve models that are more intuitive and less prone to overfitting.
\item \textit{Loss Function}: The loss function quantifies the difference between generated samples and the true data distribution, with different metrics used depending on the method. Root Mean Squared (RMS) error measures prediction accuracy, cross-entropy evaluates how similar predicted and actual distributions are, Kullback-Leibler (KL) divergence assesses how much one distribution diverges from a reference, and Wasserstein distance, while more computationally expensive, provides stability and clearer interpretation by assessing the cost of transforming one distribution into another.
\item \textit{Training, Testing, and Validation}: Training fits the model to data, validation tunes it on unseen data to avoid overfitting, and testing evaluates its generalization on a separate dataset. Ideally, the test data should be entirely unseen, but in practice, overlap often occurs, especially in chemistry. Thus, careful data curation is crucial for reliable Generative AI in computational chemistry\cite{heid2023characterizing}.
\item \textit{Regularization and Mode Collapse}:  Regularization methods, such as dropout, weight decay, and early stopping, are used to prevent overfitting by penalizing overly complex models and ensuring they generalize well to new data. Especially in generative models, care must be taken to avoid mode collapse, where the model generates limited, repetitive outputs, missing the diversity in the data. 
\item \textit{Embedding}: Embeddings are dense, low-dimensional representations that map discrete data, like words or items, into continuous vector spaces where similar items are closer together. While latent space emphasizes compressing data and extracting essential features, embedding space prioritizes capturing relationships and semantic meaning.
\item \textit{Attention}: Attention mechanisms allow models to focus on different parts of the input data when generating outputs. This is critical in transformer models and helps improve performance in tasks like natural language processing by enabling the model to weigh different input components.
\end{enumerate}

\section{Generative AI methods for computational chemistry}
Here we provide an overview of popular Generative AI methods relevant to computational chemistry. We describe the central ideas and highlight what makes these methods appealing, their limitations and new research directions towards improving them.
\subsection{Autoencoders and derived methods} 
\label{sec:autoenc}
Autoencoders have become increasingly visible in the field of computational chemistry due to their powerful ability to learn and represent complex high-dimensional molecular data as points in a low-dimensional latent space. Generally speaking, an autoencoder is a type of neural network that compresses input data into a lower-dimensional latent space and reconstructs it as accurately as possible. By sampling from this latent space, new high-dimensional molecular data can be generated, potentially discovering novel molecular configurations. Points close in the latent space correspond to similar molecules, allowing the autoencoder to explore chemical diversity (by sampling near the edges of the latent space) and classify molecular similarity (on the basis of proximity within the latent space). This strategy has been applied to classify and explore chemical space, improving similarity searches and clustering of compounds from experiments or calculations like DFT. Autoencoders are also used in reaction coordinate discovery and enhanced molecular dynamics, enabling visualization and exploration of complex, high-dimensional landscapes by reducing dimensionality while retaining key features \cite{mehdi2024enhanced}.

Autoencoders are popular in computational chemistry for their easy-to-visualize latent variables, but they are prone to misuse. A central issue is the temptation of assuming Euclidean geometry in the latent space, which can lead to uncontrolled mapping of distances between the latent and high-dimensional spaces. Additionally, latent variables may be correlated or lack physical relevance. Understanding these problems requires a deeper look into autoencoder construction, where traditionally an encoder maps input data to the latent space, and a decoder maps it back, with the goal of minimizing reconstruction loss. With this generic recipe, one can construct different types of autoencoder-inspired methods by varying the following:  

\begin{enumerate}
    \item \textit{Prior for the latent variable:} Autoencoders can minimize training loss significantly with expressive encoders and decoders, risking overfitting. To prevent this, a prior distribution is imposed on the latent variable, adding a regularization loss that keeps the latent variable close to the prior while maintaining low reconstruction loss. The choice of prior, such as a Gaussian distribution in variational autoencoders (VAEs) or a mixture of Gaussians\cite{tomczak2018vae} is an active research area. 
     \item \textit{Additional loss terms to enforce physics:} Physically meaningful latent representations can be obtained by using disentanglement-based loss terms, like in the $\beta$-VAE approach \cite{higgins2017beta}, or through dynamics-based priors \cite{wang2024latent}. In chemistry, meaningful latents can also be obtained by adding loss terms that maximize specific physical attributes, as shown in Ref. \cite{beyerle2024thermodynamically}, where the two dimensions of the latent space correspond to entropic and enthalpic degrees of freedom.
    \item \textit{Generalizing output task:} A traditional autoencoder can be extended to predict other quantities, not just reconstruct inputs. For example, using the information bottleneck framework \cite{wang2021state}, it can predict which metastable state a molecular system will visit after a time delay. This approach \cite{beyerle2022quantifying,zhao2023quantifying} closely mirrors desirable attributes of the committor function.
\end{enumerate}

\subsection{Generative adversarial networks (GANs)}
\label{sec:gan}
Generative Adversarial Networks (GANs)\cite{Goodfellow2014Generative} have gained particular attention for their ability to produce high-quality realistic images, audio, video, and chemical molecules. GANs possess two unique features:  the discriminator and the generator (Fig.~\ref{fig:genAI_framework}), which compete against each other in a zero-sum game. Here, the generator synthesizes new data while the discriminator tries to distinguish between the synthetic data and real data from a training set. Through continuous feedback, the generator progressively improves its ability to generate realistic data until the discriminator is fully deceived by the generator's newly synthesized data. This central idea underlying GANs has been improved through various variants. Conditional GANs (cGANs)\cite{Mirza2014ConditionalGA}, for instance, generate data conditioned on specific attributes, which is especially useful in chemistry for creating molecules or materials with desired properties. Wasserstein GANs (WGANs)\cite{Arjovsky2017Wasserstein} enhance training stability and mitigate issues like mode collapse by optimizing the Wasserstein distance between synthetic and real data. 

GANs have successfully demonstrated great potential for chemistry also through integration with other neural networks.
For molecular discovery, You et al.\cite{You2018Graph} have shown that combining graph neural networks with GANs, as in the Graph Convolutional Policy Network (GCPN), can effectively generate novel molecular structures by optimizing desired properties with Reinforcement Learning (Sec. \ref{sec:RL}). In addition, the idpGAN~\cite{Janson2023Direct}  model utilizes GANs by incorporating with transformer architectures (Sec. \ref{sec:language}) to generate sequence-dependent protein conformational ensembles, which can capture protein dynamics and interactions. Sidky et al.~\cite{Sidky2020Molecular} proposed latent space simulators that integrate the VAMPnet model~\cite{VAMPnets2018Andreas} with WGANs. This allows generating long synthetic MD trajectories that can accurately reproduce atomistic structures and kinetics observed in training trajectories, at a much lower computational cost. 

Despite these forays in chemistry, GANs have several limitations that restrict their applicability. These limitations include training instability, mode collapse, and a heavy dependence on large datasets. Training instability arises due to the delicate balance between the generator and discriminator, which leads to non-convergence and contributes to the issue of mode collapse. Reliance on training data makes it challenging to generate data that is out-of-distribution (OOD)\cite{Gui2023Review,Hoang2020Catastrophic}.  As a result, GANs are gradually going out of fashion for
chemical applications, and the field is shifting toward new state-of-the-art methods, such as diffusion models and reinforcement learning-based approaches, which offer solutions to some of the inherent limitations of GANs~\cite{dhariwal2021diffusion}.

\subsection{Reinforcement learning} 
\label{sec:RL}
In Reinforcement learning (RL) a proverbial agent learns to make optimal decisions by interacting with an environment to maximize cumulative rewards. RL is usually modeled as a Markov decision process, comprising states, actions, an environment, and rewards; the agent receives rewards based on its actions and the current state of the environment, with the reward signal guiding subsequent actions and state transitions. The agent employs trial and error to explore various strategies which helps it improve its actions over time. One of the most widely used applications of RL in the pharmaceutical industry is in context of computationally driven chemistry through the method REINVENT\cite{blaschke2020reinvent} and many others that have followed\cite{zhavoronkov2019deep}, which use RL often combined with other DL approaches, to generate optimized molecules consistent with user defined properties. In computational chemistry it has been used for the learning of transition states\cite{zhang2021deep}, diffusive dynamics \cite{das2021reinforcement} and sampling protein conformational dynamics\cite{kleiman2022multiagent}.
In spite of its promise, RL for chemistry continues to be plagued with a few issues fundamental to molecular systems. These include:
\begin{enumerate}
    \item \textit{Curse of dimensionality:} Molecular systems exist in high-dimensional spaces, which RL algorithms can struggle to explore and learn efficiently. Often this is dealt with an adaptive use of RL where RL is trained on some existing data set and the trained surrogate model is used to perform further exploration and data generation\cite{kleiman2022multiagent}. 
    \item \textit{Data scarcity problem:}  Important molecular events, like chemical reactions or conformational changes, occur infrequently and are hard to capture, resulting in incomplete datasets. Consequently, RL models trained on limited data may bias towards common states and miss rare but crucial phenomena.
    \item \textit{Novelty problem:} Molecular systems often have multimodal data corresponding to different metastable states. RL models often fail to generate a diverse set of outputs, leading to the mode collapse problem (Sec. \ref{sec:theory_genAI}). For instance in Ref. \cite{zhavoronkov2019deep}, the most potent design had a high structural similarity to an existing drug molecule, a problem somewhat common in the use of RL for discovery of chemical matter.
\end{enumerate} 
We conclude by highlighting how concepts from chemistry, particularly statistical mechanics, are enhancing RL. In Maximum Entropy RL (MaxEnt RL) \cite{ziebart2008maximum}, the goal is to maximize both expected reward and policy entropy, promoting stochasticity in actions. This approach improves RL algorithms' adaptability to real-world complexities. Recent work on Maximum Diffusion RL \cite{berrueta2024maximum}, based on the Principle of Maximum Caliber \cite{ghosh2020maximum}, has shown efficiency gains over traditional MaxEnt RL. The GFlowNet framework \cite{bengio2023gflownet} treats state-action trajectories as network flows, ensuring robust sampling with detailed balance and importance sampling, outperforming Markov Chain Monte Carlo methods in certain cases. The intersection of statistical mechanics and RL continues to be a promising area for developing innovative RL methods.
\subsection{Flow based methods} 
\label{sec:flowmethods}
Having discussed recent RL variants inspired by statistical physics in the last subsection, we now turn to models rooted in these principles from their inception \cite{sohl2015deep}. Flow models aim to sample from an inaccessible target probability distribution using limited available samples -- an empirical dataset $\mathcal{D}$. Physics-inspired methods like simulated tempering \cite{SimulatedTempering} and annealed importance sampling \cite{AnnealedImportanceSampling} transform samples from the simple prior into samples from the target by constructing a bridge between the distributions. However, constructing the bridge relies on modifying the energy functions of the target and prior distributions. Generative flow models use neural networks to bridge distributions without requiring access to the energy function.

Flow-based models transform a simple prior distribution $q(\mathbf{z})$ into a more complex target distribution $p(\mathbf{x})$ through a series of learnable mappings (Fig.~\ref{fig:genAI_framework}). Normalizing flows is a popular framework for sampling from $p(\mathbf{x})$. A normalizing flow model parameterizes an invertible transformation $\mathcal{M}$. The transformation acts as a change of variables that deforms the prior into the target distribution so that sampling from the prior and applying $\mathcal{M}$ produces a sample from the target distribution, i.e. $\mathbf{z} = \mathcal{M}(\mathbf{x})$. The change in probability associated with the change of coordinates is given by the identity

\begin{equation}
    \label{eq:change of measure}
    \log p(\mathbf{x}) = \log q(\mathbf{z}) + \log \left| \mathrm{det} J_\mathcal{M}(\mathbf{x}) \right|.
\end{equation}
The identity in eq. \ref{eq:change of measure} reweights samples between the prior and target distributions via the Jacobian determinant of $\mathcal{M}$. The optimal change of variables $\mathcal{M}$ maximizes the likelihood for samples in $\mathcal{D}$ and can be obtained by training a neural network $\mathcal{M}_\theta$ to maximize $\log p(\mathbf{x})$.

However, computing the determinant is a (usually) prohibitively expensive operation that scales as $\mathcal{O}(d^3)$ in the general case, where $d$ is the number of components of $\mathbf{x}$. Normalizing flows address this by employing architectures that result in structured Jacobians (for example, alternatingly upper- and lower-triangular\cite{RealNVP}) for which the determinant computation is faster. However, the computationally tractable determinant results in reduced expressivity, and much recent effort has been devoted to addressing this tradeoff \cite{neural-splines, stochastic-normalizing-flows}.


Diffusion models\cite{ddpm} are generative algorithms that do not require access to the Jacobian determinant during training. Diffusion models learn to invert an Ornstein-Uhlenbeck (OU) diffusion process. The diffusion process generates a probability flow $p(\mathbf{x},t)$ that transports the target distribution $p(\mathbf{x},t=0)$ to the prior $q(\mathbf{z}) \equiv p(\mathbf{x},t=1)$\cite{sgm-sde}. Diffusion models are made possible by a theorem stating that the drifts of the forward- and reverse-time diffusion processes differ only by the score $\nabla_\mathbf{x} \log p(\mathbf{x},t)$ \cite{Anderson1982}. Similar to normalizing flows, a neural network is trained to estimate the score from realizations of the forward process. Once parameterized, the score network can be used to simulate the time-reversed diffusion process, thereby generating samples from the target distribution.

The ideas behind diffusion models are deeply rooted in statistical physics. The OU diffusion process is naturally studied using techniques from non-equilibrium thermodynamics\cite{sohl2015deep}, and parametrizing the gradient of a potential (the score) rather than the potential itself is similar in spirit to force-matching in coarse-grained models \cite{Jin2022}. Looking forward, tools from statistical physics will be central to understanding how diffusion models operate. For example, using techniques from spin-glass theory it was found that the dynamics of the diffusion processes can be divided into distinct regimes distinguishing generalization and memorization of the training data \cite{dynamical-regimes-mezard}. Recent exciting advancements such as flow-matching\cite{flow-matching, CFM} and stochastic interpolants\cite{stochastic-interpolants} further introduce ideas from optimal transport to model arbitrary diffusion processes.

\subsection{Recurrent neural networks and large language models} 
\label{sec:language}
Recurrent neural networks (RNNs), like Long Short-Term Memory (LSTM) networks, and transformer-based architectures \cite{hochreiter1997long,vaswani2017attention}, have made significant strides in natural language processing, speech recognition, and computational chemistry, notably in protein structure prediction with AlphaFold2. These models excel in handling sequential data, whether predicting the next value in a time series, generating sequences, or classifying entire sequences. LSTM networks use memory cells and gating mechanisms to maintain long-term dependencies, while transformers, such as those in large language models (LLMs), use self-attention mechanisms to capture complex relationships, processing sequences more efficiently and effectively. A typical LLM can contain an encoder and a decoder (Fig. \ref{fig:genAI_framework}) both composed of multiple transformer layers, each leveraging self-attention to understand the input context. The attention mechanism is key in aligning and focusing on the most relevant parts of the sequence during both encoding and decoding, enabling the model to generate coherent and contextually appropriate outputs.

These models are particularly valuable for chemistry, where many processes are non-Markovian and require long-term dependencies for accurate predictions, such as in reaction prediction and molecular dynamics simulations. However, the success of LLMs in other domains doesn't easily transfer to chemistry due to their limitations in extrapolating beyond the training data and not accounting for hidden biases \cite{bender2021dangers}. This is critical in a field where novel molecules and reactions often lie outside previously explored chemical spaces. To address this, specialized approaches are needed, such as the one proposed in \cite{tsai2022path}, which integrates statistical physics into LSTM training using a path sampling approach based on the Principle of Maximum Caliber \cite{ghosh2020maximum}. 

\section{Selected applications}

\subsection{\textit{Ab initio} quantum chemistry and coarse-grained force fields}
\label{sec:quantum}
In quantum chemistry, deep neural networks are used to achieve high quantum-level accuracy while reducing computational costs. 
For instance, AI algorithms have been developed to solve electronic Schr\"odinger equations for ground and excited states, reducing the computational complexity from $O(N^7)$ to $O(N^4)$ \cite{hermann2020deep,david2024accurate,ratcliff2017challenges}. As for MD simulations,
machine learning force fields  allow \textit{ab initio} quantum-quality calculations to approach the speed of classical MD simulations \cite{tiwary2024modeling}. Another approach directly using Generative AI is applying coarse-grained models for macromolecules, which effectively reduce the number of atoms in MD simulations \cite{majewski2023machine}. Ab initio MD is computationally expensive, but MLFFs can estimate energies and forces from atomic configurations without electron calculations \cite{noe2020machine}, enabling quantum-quality simulations of hundreds of atoms over nanoseconds. MLFFs have been used to study liquid dielectric constants \cite{zhang2023why}, phase behaviors of water \cite{gartner2022liquid}, proton transfer for energy materials \cite{quaranta2017proton}, and prebiotic chemical reactions \cite{benayad2024prebiotic}. Beyond MLFFs, diffusion models can generate molecular structures by sampling the Boltzmann distribution, optimizing geometry without force and energy calculations \cite{rothchild2024investigating}, speeding up ground-state searches. Additionally, ML can model charge density from structures \cite{pope2023towards}, which, while not always needed for MD, provides the missing charge density in MLFFs, expanding AI applications to complex quantum chemistry problems like predicting vibrational spectra \cite{zhang2020prb}. Developing MLFFs that can generalize beyond training data remains an area of concern and active research interest \cite{zhai2023short,benayad2024prebiotic}

While MLFFs maintain quantum-level accuracy in systems typically studied with classical MD, CG models simplify simulations by representing larger systems with reduced atomic detail. An open challenge in CG models is backmapping to all-atom configurations. One approach uses an auto-encoder architecture, where the encoder learns the CG model, and the decoder backmaps it to an ensemble's average structure \cite{wang2019coarse}. Denoising Diffusion Probabilistic Models can also generate atom-resolution structures \cite{jones2023diamondback}. Additionally, learning the score term of a denoising diffusion process can approximate CG force fields and generate equilibrium Boltzmann distributions around local energy minima \cite{arts2023two}.

\subsection{Protein structure and conformation prediction} 
Understanding a protein's structure, both in native and non-native states, is crucial for determining its function, stability, and interactions. Traditionally, structure prediction relied on experimental techniques like X-ray diffraction (XRD), nuclear magnetic resonance (NMR), and cryo-electron microscopy (Cryo-EM) \cite{Goh2016Computational}. The success of AI-driven approaches like AlphaFold2 (AF2) and RoseTTAFold, which can now predict crystal-like protein structures, is largely due to the availability of such high-quality experimental data deposited in the Protein Data Bank (PDB) \cite{Jumper2021Highly,Baek2021Accurate}. AF2 uses co-evolutionary information from multiple sequence alignments (MSAs) to predict structures from amino acid sequences, while RoseTTAFold enhances accuracy by integrating evolutionary information with 3D coordinate refinement \cite{Baek2021Accurate}. Recently, AF3 and RFDiffusion have further advanced predictions by incorporating diffusion-based models capable of handling complex structures \cite{Abramson2024Accurate,Watson2023Denovo}.

While AI methods have transformed biomolecular structure prediction, they remain limited by the quality of their training datasets and struggle to predict metastable non-native structures or the effects of point mutations \cite{Buel2022Can}. To explore hidden conformations, approaches like MSA perturbation have been developed, including reduced-MSA \cite{Alamo2022Sampling}, AF2Cluster \cite{Wayment_Steele2024Predicting}, and SPEECH-AF \cite{Stein2022Sampling}, all providing structural diversity. However, proteins are dynamic systems with fluctuations that depend precisely on thermodynamic parameters such as temperature, pressure, and chemical potential \cite{Henzler2007Dynamic,Bowman2024AlphaFold}. To capture this conformational diversity, methods like AF2RAVE \cite{Vani2023AF2RAVE} and AF2 integrated with MSM \cite{Meller2023Accelerating} generate hypothetical structures through MSA perturbation and rank them according to their Boltzmann weight via MD simulations. AF2RAVE, for example, uses reduced-MSA for initial predictions, clusters them with the State Predictive Information Bottleneck approach \cite{wang2021state}, and ranks them using enhanced sampling MD simulations. This method has been effective in capturing conformational changes, such as DFG-in and -out conformations of tyrosine-protein kinases, improving accuracy in downstream docking \cite{Vani2024Exploring,Gu2024Empowering}.

In some cases, MD simulations are used as training data for AI-driven methods. The Boltzmann generator \cite{Noe2019Boltzmann} combines MD simulations with normalizing flows to generate Boltzmann-weighted molecular conformations, potentially applicable to proteins. AlphaFlow \cite{jing2024alphafold} integrates AlphaFold with flow matching \cite{lipman2023flow} to produce diverse protein conformations, training on MD simulations from the ATLAS dataset \cite{Vander2023ATLAS} to enhance diversity and capture fast mode dynamics. However, it struggles with slow, emergent properties, limiting its ability to model long-term protein behavior. Extending training with longer MD simulations could address this but raises questions about the added value of Generative AI. Two recent approaches, Distributional Graphormer (DiG) \cite{Zheng2024Predicting} and NeuralPLexer \cite{qiao2024state}, combine Generative AI with MD to predict conformational distributions. DiG uses a diffusion algorithm within a graphformer architecture to generate protein conformations from a primary sequence, with MD simulations as training data, while NeuralPLexer employs a diffusion-based model to predict state-specific protein-ligand complex structures, capturing dynamic conformational changes by integrating biophysical constraints.

As demonstrated in these examples, Generative AI and its further incorporation with MD simulations have proven effective in accurately predicting protein structures and exploring protein conformations, bringing out the best of both AI and MD. We expect this will remain an area of active research with the continued development of more robust and accurate methods. That said, we urge caution in training Generative AI on synthetic MD data. It's essential these models respect physical laws, as deviations could lead to non-physical predictions. Like with DALL-E and ChatGPT, deepfakes could proliferate, leading to unreliable outcomes and unreal chemistry.

\subsection{RNA structure prediction} 
\label{sec:rna}
RNAs are an emerging frontier in medicinal chemistry\cite{hargrove2021targeting}, yet despite their growing interest, experimentally solved RNA structures remain scarce\cite{szikszai2024rna3db}. As a result, computational methods have become indispensable for modeling RNA tertiary structures. Among all methods, physics-based computational methods -- like minimum free energy secondary structure prediction algorithms\cite{nussinov,zuker1989,mfold2003,xu2014vfold} and energy-based simulations\cite{parisien2008mc-fold,sripakdeevong2011enumerative,watkins2020farfar2} -- are especially prevalent. These methods model energetic and entropic contributions to RNA structure formation and are a starting point for successful generative approaches. An intuitive approach to integrating AI with physics-based methods is using a neural network to predict geometries that are later refined with an energy function \cite{pearce2022novo,wang2023trrosettarna,li2023integrating}. While these methods have the potential to generalize well across diverse RNA sequences, their accuracy currently lags behind methods that make use of large databases\cite{griffiths2003-rfam,rnacentral2019-RNAcentral} of RNA sequences. Future advancements in this space might be more improved energy\cite{stasiewicz2019qrnas,briq2021} or scoring functions\cite{wang20153drnascore,townshend2021geometric} and conformational sampling algorithms \cite{li2023rnajp,chen2023rna-alchemy2}.

Language models for RNA were developed following their widespread success in machine translation and text generation\cite{akiyama2022-RNAbert,chen2022-RNA-FM,wang2023-UNI-RNA,chen2023-spliceBERT,boyd2023-ATOM-1,chu2024-UTR-LLM}. These models parameterize embeddings that can serve as inputs for other networks fine-tuned for specific tasks like structure or function prediction. AF3's MSA module and Pairformer\cite{Abramson2024Accurate} learn embeddings from multiple sequence alignments, while foundation models like RNA-FM\cite{chen2022-RNA-FM} or ATOM-1\cite{boyd2023-ATOM-1} learn from single sequences or chemical mapping data. The premier application of RNA language models to structure prediction is as a module within a larger multi-component AI system. 

Recent years have seen the emergence of large-scale pre-trained Generative AI models for RNA structure prediction\cite{shen2022-e2efold-rhofold,baek2022accurate,kagaya2023-nufold,krishna2024generalized,Abramson2024Accurate}. These are neural networks composed of different ``blocks'' or ``modules'' that learn specific aspects of biomolecular structure. Together, the modules predict tertiary structures from sequence and template information. AF3\cite{Abramson2024Accurate} and RoseTTAFold-2NA\cite{baek2022accurate} (RF2NA) are state-of-the-art in RNA tertiary structure prediction, though the accuracy of predicted RNA structures does not yet rival that of proteins structures\cite{das2023-CASP15}. Still, both methods aim to predict one structure per sequence -- a modeling paradigm increasingly challenged by advances in structural biology.

The field of structural biology is undergoing a paradigm shift, moving away from the historically prevalent native-centric view of biomolecular structure\cite{henzler2007-protein-ensemble,bonilla2022-rna-cryo-ensemble,ken2023-rna-ensemble-excited,bonilla2024-rna-ensemble}. A single structure may be insufficient to account for biomolecular function; the appropriate description is instead an {\it ensemble} -- a set of representative Boltzmann-weighted conformers. Currently, the goal of most Generative AI methods for RNA structure prediction is to predict one native structure per sequence\cite{Bowman2024AlphaFold}. To fully embrace this paradigm shift future generative methods should aim to predict Boltzmann-weighted structural ensembles.

To complicate matters further, the ensemble is not static: it responds to environmental factors like temperature\cite{kortmann2012-rna-thermometer}, ions\cite{brion1997-rna-ions}, or small molecules\cite{orlovsky2019-hiv-tar-small-molecule}. While AF3 can model the environment explicitly to some extent, it is perhaps more feasible to model the environment implicitly. Thermodynamic Maps\cite{herron2023inferring} (TMs) follow a generative framework that infers the structural ensemble's dependence on environmental conditions. TMs are diffusion models where thermodynamic parameters (e.g. temperature) are implicitly represented within the Langevin dynamics of the diffusion process. The central idea is to map molecular conformations onto those of a simple, idealized system where the effect of the environment is straightforward to account for. Though still in its early stages, TMs have demonstrated the ability to predict the temperature dependence of RNA structural ensembles.

\begin{figure}
\centering
\includegraphics[width=.85 \linewidth]{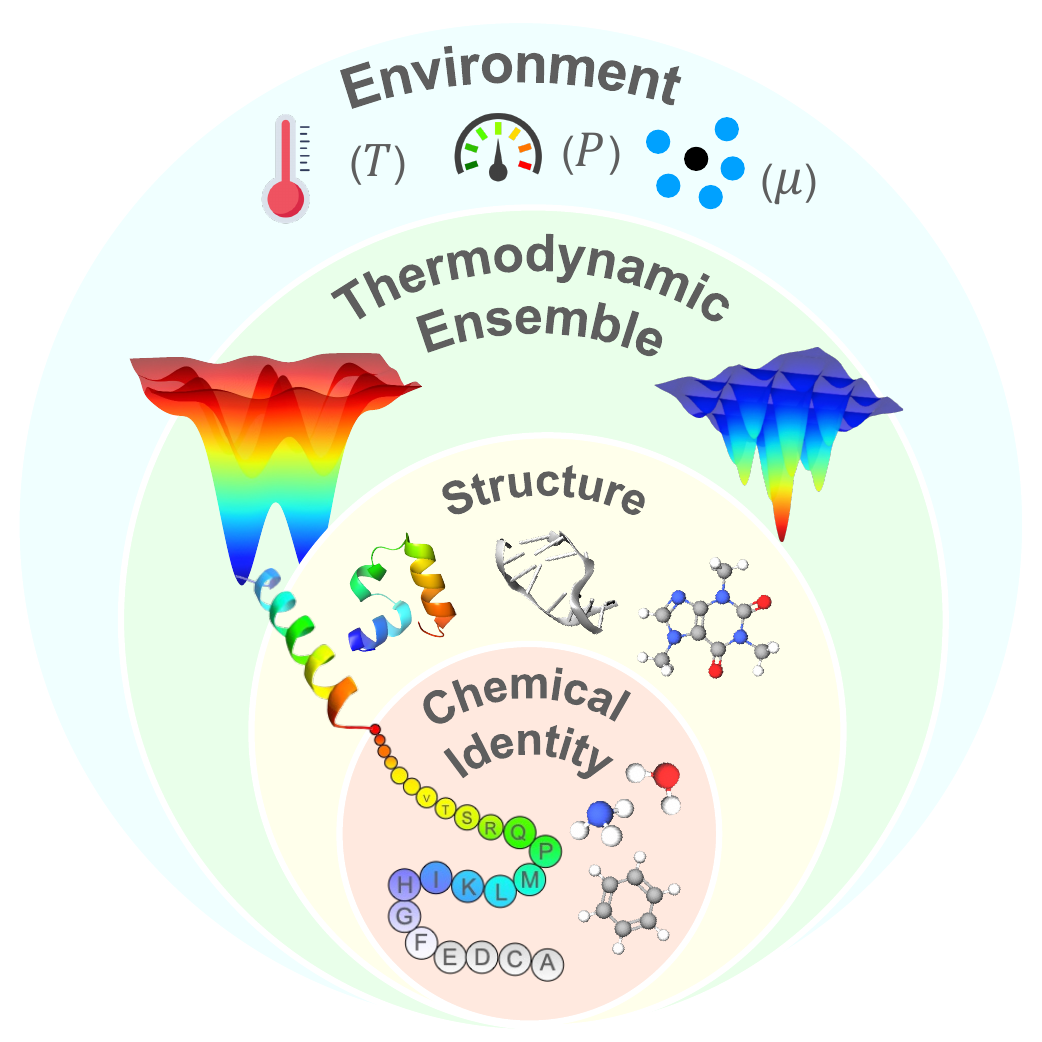}
\caption{ \textbf{Desirables for predicting emergent phenomena in chemistry with Generative AI.} The goal for Generative AI, molecular simulations, and computational chemistry is to start from the chemical identity—whether sequence or composition—and accurately predict function while considering the relevant environmental conditions. To achieve this, the intermediate rungs of structure, thermodynamic ensemble, and environment must be accounted for, where the environment can be quantified through parameters such as the Temperature $T$, pressure $P$ and chemical potential $\mu$. As we move up each level, modeling becomes more challenging due to the nuances of fluctuations governed by laws of equilibrium and non-equilibrium statistical mechanics. This will require Generative AI models deeply grounded in statistical mechanics, with precise priors to account for complex interactions, dynamics, and environmental effects.}
\label{fig:genai_desirable}
\end{figure}

\section{Desirables from Generative AI for chemistry} 
We believe the ultimate predictive power of any tool—whether theory, molecular simulations, or Generative AI—lies in starting from chemical identity and accurately predicting function while rigorously accounting for environmental conditions. However, function is not an inherent property of a given sequence or a chemical formula; rather, it is an emergent property that arises from the dynamic interactions and feedback loops across multiple scales. Achieving this requires navigating through increasing complexity—structure, thermodynamic ensemble, and environment (Fig. \ref{fig:genai_desirable})—where modeling challenges intensify due to intricate fluctuations governed by equilibrium and non-equilibrium statistical mechanics. While current Generative AI methods like AlphaFold2 predict the most stable structure given chemical identity, much more remains to be desired. This section outlines key attributes for advancing Generative AI in this direction.
\begin{enumerate}
\item \textit{Chemistry and AI, Not AI for Chemistry}:
It is indubitable that AI's ability to handle large, complex datasets and uncover hidden structures makes it especially valuable in computational chemistry. At the same time, integrating chemistry, particularly statistical mechanics, quantum mechanics, and thermodynamics, with AI creates a powerful synergy. These fields provide essential priors, frameworks for hypothesis testing, and tools for extrapolation that enhance Generative AI's effectiveness in chemistry. 

\item \textit{Interpretability and Reliability Testing}:
The interpretability and reliability of AI models are paramount in chemistry. Current internal confidence measures, such as AlphaFold’s pLDDT scores, have shown limitations in providing reliable assessments of model predictions\cite{Buel2022Can}. A focus on developing more robust interpretability frameworks and reliability testing is necessary to ensure that AI models not only generate accurate predictions but also provide meaningful insights into their confidence levels and potential errors.

\item \textit{Out-of-Distribution Generalization and Efficacy in Data-Sparse Regimes}:
AI excels at smart interpolation within well-covered data regions, but the real challenge is generalizing to out-of-distribution data to reduce hallucinations and spurious predictions. In chemistry, where data is often sparse, Generative AI must be effective with limited data, transferring learned knowledge across diverse chemical types to ensure models remain robust, versatile, and reliable in scientific discovery.

\item \textit{Rethinking Data - More data is not always better data}:
In chemistry, the typical scaling laws seen in large language models—where more data improves test loss—may not hold true. More data doesn't always enhance performance and can sometimes obscure meaningful insights. For instance, in an MD trajectory trapped in metastable states, adding more data might amplify noise rather than provide useful information about rare transitions. This underscores the need to rethink data handling in AI for chemistry, ensuring that additional data is genuinely useful and doesn't obscure key physical insights. Moreover, traditional AI methods of splitting data into training, testing, and validation sets can be problematic due to data leakage and challenges in quantifying overlap.

\item \textit{Emergent Phenomena and correct coupling to environmental variables}:

The power and thrill of MD and computational chemistry lie in predicting new chemistry and physics that emerge naturally when simulating a large system for a long time—phenomena that couldn't be guessed from a simple force field or Hamiltonian. Emergent behavior, sensitive to control parameters and environmental variables, often arises in such simulations, closely aligning with theoretical predictions. However, Generative AI often struggles to produce novel physics or chemistry beyond its training data, with emergent phenomena sometimes being artifacts. Most AI models also fail to account for environmental conditions, limiting their predictive power in new scenarios. Hybrid approaches like AF2RAVE \cite{Vani2023AF2RAVE} and Thermodynamic Maps \cite{herron2023inferring} show promise in this context by integrating AI with physical principles.
\end{enumerate}

\section{Critical assessment and outlook}
Generative AI has made impressive strides in computational chemistry, particularly in force field development, structure prediction, and accelerated molecular simulations, showing its potential to tackle complex chemical challenges. However, significant obstacles remain before AI can fully integrate into the molecular simulation toolbox. The ultimate goal of any simulation method or theory is to reliably predict chemical function directly from chemical identity—a dream yet unrealized. We argue that the same aspirations should be applied to Generative AI for physical sciences. By integrating chemistry, particularly statistical mechanics, into AI models, considering the roles of structural and dynamical ensembles with precise fluctuations, and accounting for environmental influences, this goal may be achievable. While chemistry has much to gain from Generative AI, it also has much to teach it. Grounding AI in chemistry's principles can create more accurate, adaptable, and interpretable models. This integration could transform AI into a powerful tool for predicting novel emergent phenomena, driving discoveries, and deepening our understanding of chemical processes.

\acknow{This work was supported by NIH/NIGMS under award number R35GM142719. We thank UMD HPC’s Zaratan and NSF ACCESS (project CHE180027P) for computational resources. P.T. is an investigator at the University of Maryland-Institute for Health Computing, which is supported by funding from Montgomery County, Maryland and The University of Maryland Strategic Partnership: MPowering the State, a formal collaboration between the University of Maryland, College Park and the University of Maryland, Baltimore.}

\showacknow 


\end{document}